\documentclass{webofc}

\usepackage[varg]{txfonts}   
\usepackage{hyperref}
\usepackage{url}
\hypersetup{colorlinks=true,citecolor=blue,urlcolor=blue,linkcolor=blue}
\begin{document}
\title{Disoriented isospin condensates in heavy-ion collisions}
%
%

\author{\firstname{Joseph} \lastname{Kapusta}\inst{1}\fnsep\thanks{\email{kapusta@umn.edu}} \and
        \firstname{Scott} \lastname{Pratt}\inst{2}\fnsep\thanks{\email{prattsc@msu.edu}} \and
        \firstname{Mayank} \lastname{Singh}\inst{1,3}\fnsep\thanks{\email{mayank.singh@vanderbilt.edu}}
}

\institute{School of Physics and Astronomy, University of Minnesota, Minneapolis, MN 55455, USA
\and
           Department of Physics and Astronomy and Facility for Rare Isotope Beams, Michigan State University, East Lansing, MI 48824, USA
\and
           Department of Physics and Astronomy, Vanderbilt University, Nashville, TN 37240, USA
          }

\abstract{Anomalous neutral to charged kaon correlations measured by the ALICE collaboration have defied usual explanations. We propose that the large fluctuations could arise because of a disoriented isospin condensate where there is an imbalance between up and down condensates at the time kaons hadronize. This could happen in heavy-ion collisions when the quark condensate re-forms as the system cools and the approximate chiral symmetry of QCD is broken. Within the linear sigma model, we show that the energy cost of forming a disoriented isospin condensate is small making it very plausible.
}
\maketitle
\section{Introduction}

The ALICE collaboration at the Large Hadron Collider (LHC) has reported anomalous fluctuations between charged and neutral kaons produced in Pb-Pb collisions at $\sqrt{s_{NN}} = 2.76$ TeV \cite{ALICE:2021fpb}. The fluctuations were significantly larger than what was predicted by conventional heavy-ion simulators like AMPT, HIJING and EPOS-LHC \cite{Nayak:2019qzd}. The measured quantity was $\nu-$dynamic which is defined as \cite{Gavin:2001uk}

\begin{equation}
    \nu_{\rm dyn}(K_\pm,K_S) = \frac{\langle N_\pm(N_\pm -1)\rangle}{\langle N_\pm\rangle^2} + \frac{\langle N_S(N_S -1)\rangle}{\langle N_S\rangle^2} - 2\frac{\langle N_\pm N_S\rangle}{\langle N_\pm\rangle\langle N_S\rangle}.
\end{equation}
Here, $N_\pm$ denotes the number of charged kaons while $N_S$ is the number of neutral K-short kaons. The averaging is done over many events.

The ALICE collaboration found that the third term, the covariance between charged and neutral kaons was large and negative. This indicates that there were regions with an excess of either charged or neutral kaons relative to the other. There are a few processes like charge conservation, resonance decays or Bose symmetrization which can cause these local fluctuations. But the measured $\nu_{\rm dyn}$ exceeded the expectations from these effects.

The above-mentioned effects tend to be local. Yet, the measured anomaly extended to over one unit in rapidity, indicating that these fluctuations are correlated over a long range. The fluctuations from known effects tend to be proportional to inverse multiplicity. When $\nu_{\rm dyn}$ was scaled with $\alpha = 1/\langle N_\pm\rangle + 1/\langle N_S\rangle \approx 6/N_K^{\rm tot}$, the anomaly was larger for events with more particles. Here $N_K^{\rm tot}$ is the total number of kaons and is expected to be proportional to the total multiplicity. All this strongly suggests that the source of these correlations is unlikely to be local.

The disoriented isospin condensates (DIC) can explain the data. It will show a similar anomaly when $\nu_{\rm dyn}$ is measured for a pair of particles one of which is rich in up quarks and the other in down quarks.

\begin{figure}
    \centering
    \includegraphics[width=0.45\linewidth]{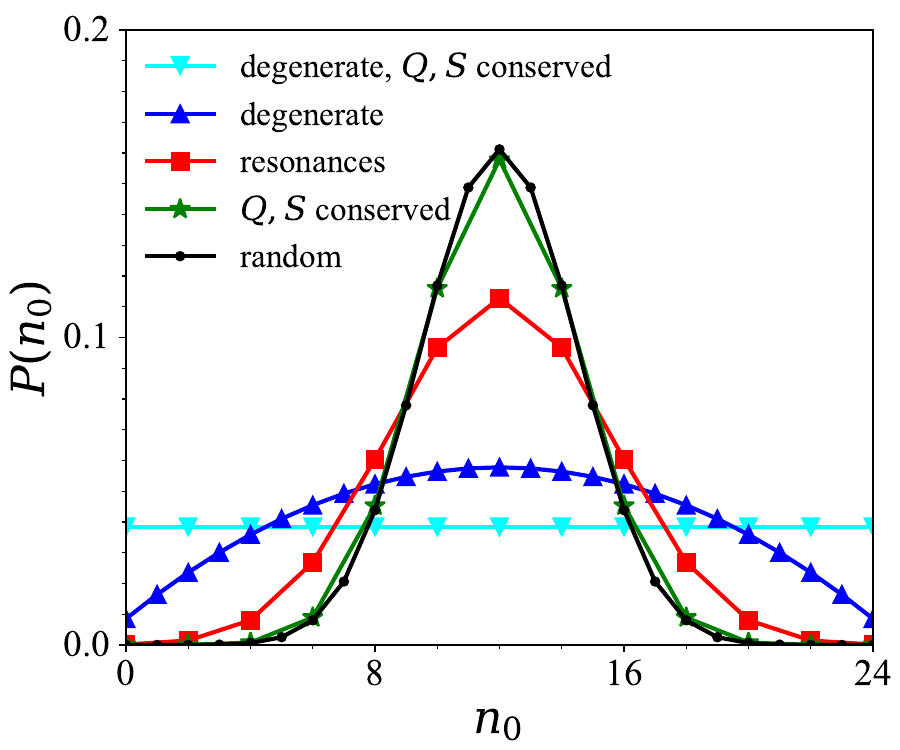}
    \includegraphics[width=0.45\linewidth]{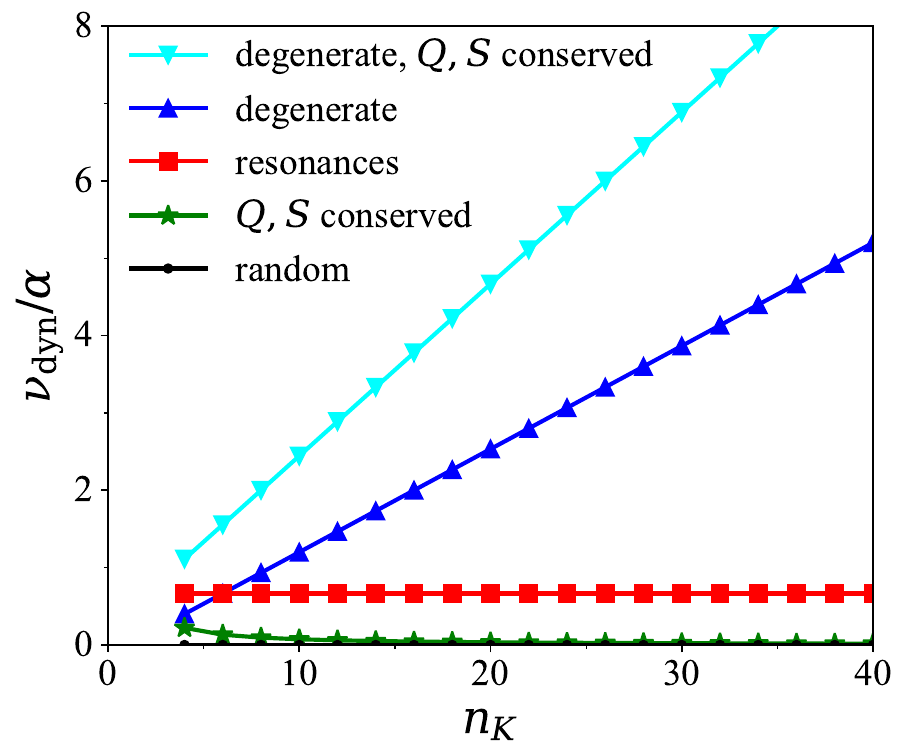}
    \caption{Probability density of $n_0$ kaons being neutral in a box of 24 kaons with different distributions (left). With inclusion of charge conservation, only even values of $n_0$ are allowed and $P(n_0)$ has been scaled by half. The $\nu_{\rm dyn}/\alpha$ for a single domain with $n_K$ kaons (right).}
    \label{fig1}
\end{figure}

\section{Domains of flat neutral kaon fraction}
The $\nu_{\rm dyn}/\alpha$ can be calculated for simple kaon systems with different distributions as shown in figure \ref{fig1} \cite{Kapusta:2022ovq}. When the kaons are distributed randomly into its four types, then $\nu_{\rm dyn}$ is identically zero. Charge and strangeness conservation gives it a non-zero value but that rapidly decreases to zero with increasing particle numbers. Resonance decays give a multiplicity independent $\nu_{\rm dyn}/\alpha$. Within these simple systems, only a box of degenerate kaons can reproduce the experimental trend of $\nu_{\rm dyn}/\alpha$ increasing with multiplicity. While degenerate kaons are extremely unlikely in a heavy-ion collision it is important to note that a flat or near-flat neutral kaon fraction distribution has the right qualitative trend as compared to data. We show in the next section that DIC can produce a similar neutral kaon fraction distribution.

In a real experimental event, the kaon domains are not isolated. There can be causally disconnected volumes of kaons with different neutral kaon fractions. They have to be folded with each other and with thermal kaons which are produced with a nearly random distribution between different types. If the number of domains $N_d$ is greater than 2, it can be shown that \cite{Gavin:2001uk}

\begin{equation}\label{eq1}
    \nu_{\rm dyn} = 4 \beta_K \left( \frac{\beta_K}{3N_d} - \frac{1}{N_K^{\rm tot}} \right).
\end{equation}
Here $N_d$ is the number of domains and $\beta_K$ is the fraction of kaons coming from these domains. Assuming that the number of domains is proportional to the size of the system and that $\beta_K = \epsilon_\zeta V_d/m_K N_K^{\rm tot}$ where $\epsilon_\zeta$ is the energy density of domain formation, $V_d$ is the total volume of all domains put together and $m_K$ is the mass of a kaon, we can do a two parameter fit to the ALICE data \cite{Kapusta:2022ovq}.

\begin{figure}
    \centering
    \includegraphics[width=0.5\linewidth]{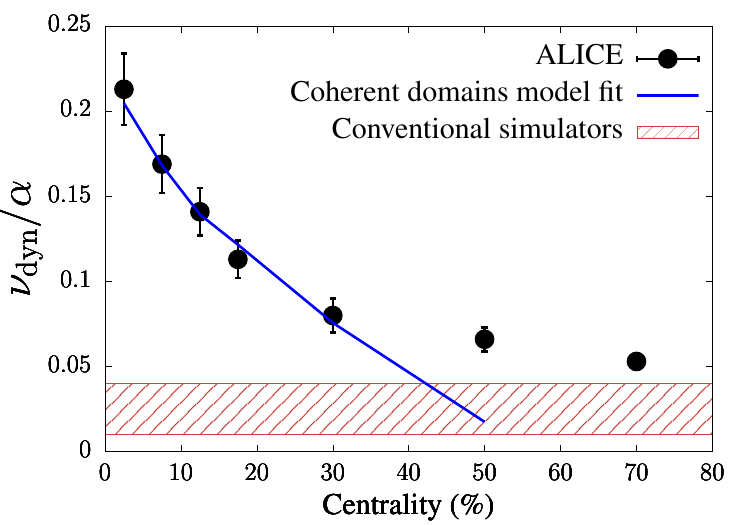}
    \caption{A model with domains of flat neutral kaon fraction can explain the data. The experimental data is from ref. \cite{ALICE:2021fpb}.}
    \label{fig2}
\end{figure}

Figure \ref{fig2} shows that a model with domains of flat neutral kaon fraction can explain the data. Only the five central points are fit as the number of domains fall below two for the last two data points. In this regime eq. \ref{eq1} is not valid. Also, note that the discrepancy between the data and the conventional models is small for larger centralities.

\section{Disoriented isospin condensates}

When the quark condensates re-form as the hot quark gluon plasma cools in heavy-ion collisions, they ultimately attain the vacuum equilibrium value where $\langle u\Bar{u}\rangle = \langle d\Bar{d}\rangle$. But they could have fluctuations over small volumes at finite temperatures as they re-form. These fluctuations could result in $\langle u\Bar{u}\rangle \neq \langle d\Bar{d}\rangle$ \cite{Kapusta:2023xrw}. This can happen by combination between the isosinglet $\langle u\Bar{u}\rangle + \langle d\Bar{d}\rangle$ and isotriplet $\langle u\Bar{u}\rangle - \langle d\Bar{d}\rangle$ states. So the light quark condensate can range from all $\langle u\Bar{u}\rangle$ to all $\langle d\Bar{d}\rangle$. If the hadronization happens when DICs are present, they will leave their imprint on the final hadrons. If there is an excess of $\langle u\Bar{u}\rangle$, then their combination with strange quarks and anti-quarks will lead to an excess of charged kaons over neutral kaons. The opposite will happen if $\langle d\Bar{d}\rangle > \langle u\Bar{u}\rangle$. So, if there are domains of DIC, they will lead to domains of flat neutral kaon fraction which will explain the ALICE data.

But how plausible is it to form DICs? We can calculate the statistical probability of condensate fluctuations using the 2+1 flavor linear sigma model \cite{Schaffner-Bielich:1998mra}. The scalar fields can be related to the condensates as $\sigma_u = - \langle \bar{u}u \rangle/\sqrt{2}c'$, $\sigma_d = - \langle \bar{d}d \rangle/\sqrt{2}c'$, and $\zeta = - \langle \bar{s}s \rangle/\sqrt{2}c'$. Here $c'$ is a constant. The up and down scalars can be related to $\sigma$ as $\sigma_u = \sigma \cos\theta$ and $\sigma_d = \sigma \sin\theta$. Then $\theta$ ranges from 0 to $\pi/2$. The stable equilibrium value is obtained when $\theta = \pi/4$.

\begin{figure}
    \centering
    \includegraphics[width=0.45\linewidth]{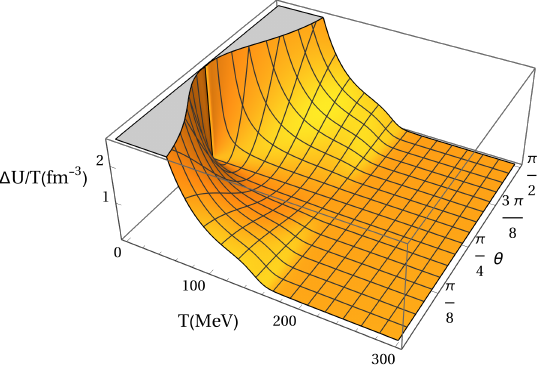}
    \includegraphics[width=0.45\linewidth]{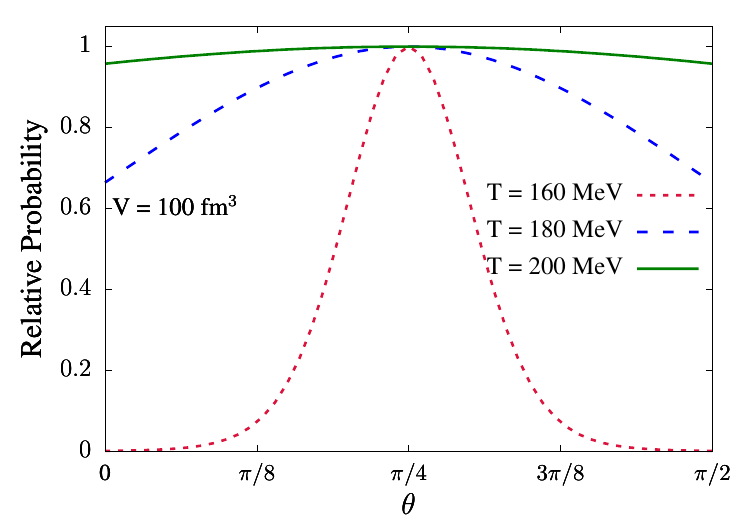}
    \caption{Energy cost of forming a DIC (left) and the relative probability for different values of $\theta$ for a domain of $V = 100$ fm$^3$.}
    \label{fig3}
\end{figure}

We can use lattice calculations to obtain the temperature dependence of quark condensates \cite{Bazavov:2011nk}. The energy cost for $\theta \neq \pi/4$ at temperature T is given as
\begin{equation}
    \Delta U(\theta, T) = \frac{1}{2}\lambda \left[1-\sin^2(2\theta) \right]\sigma^4+c(T) \left[1-\sin(2\theta)\right]\sigma^2\zeta + f_\pi m_\pi^2
\left[
1-\frac{\cos\theta+\sin\theta}{\sqrt{2}}
\right]\sigma.
\end{equation}
The parameters are fixed to reproduce the known vacuum masses of mesons and are described in \cite{Kapusta:2022ovq,Kapusta:2023xrw}. The energy cost as a function of $\theta$ and temperature is plotted in figure \ref{fig3} (left). At large temperatures, when the condensates have melted, there is negligible cost to $\theta$ fluctuations. At low temperatures, it becomes prohibitively expensive to form a DIC and $\theta$ is restricted to $\pi/4$. For intermediate temperatures, importantly around hadronization temperature, there is a finite chance of $\theta$ fluctuations.

We can also calculate the relative probability of having a particular $\theta$ value at different temperatures. For a domain of volume $V$, the relative probability is given by
\begin{equation}
    P(\theta) = e^{-\Delta U (\theta, T)V/T}.
\end{equation}
The relative probability is shown for $V = 100$ fm$^3$ in figure \ref{fig3} (right). This will translate into similar plots for a neutral kaon fraction distribution relevant for the ALICE measurement.

\section{Conclusions}

The neutral to charged kaon fluctuations measured by ALICE collaboration cannot be accounted by known effects. They can be explained if domains of flat or near flat neutral kaon fractions exist in the system. Disoriented isospin condensates can form such domains. Thermodynamic analysis tells us that the energy cost of forming DIC is really small and hence they are quite plausible. 

Recently it was shown that DICs go to their equilibrium values on time scales much shorter than those relevant for heavy-ion collisions \cite{Chabowski:2024smx}. A full theory of DIC integrated with the hydrodynamic model of heavy-ion collisions can help us make predictions and constrain the physics of chiral symmetry restoration in these systems. This can be informed by lattice QCD calculations.

\section*{Acknowledgements}
This work was supported by the U.S. Department of Energy Grant Nos. DE-FG02-87ER40328 (JK and MS), DE-FG02-03ER41259 (SP) and DE-SC-0024347 (MS).
%
%

\end{document}